\documentstyle[12pt]{article}

\global\arraycolsep=1pt
\newcommand{\beq}{\begin{equation}}
\newcommand{\eeq}{\end{equation}}
\newcommand{\beqa}{\begin{eqnarray}}
\newcommand{\eeqa}{\end{eqnarray}}
\newcommand{\CR}{\nonumber \\}

\renewcommand{\theequation}{\thesection.\arabic{equation}}
\renewcommand{\thefootnote}{\fnsymbol{footnote}}

\begin{document}

\begin{titlepage}
\begin{flushright}
{August, 2002} \\
{\tt hep-th/0208139}
\end{flushright}
\vspace{0.5cm}
\begin{center}
{\Large \bf
Harmonic Forms and  Deformation of ALC metrics with $Spin(7)$ holonomy
}%
\lineskip .75em
\vskip1.0cm
{\large Hiroaki Kanno\footnote{e-mail: kanno@math.nagoya-u.ac.jp}}
\vskip 1.0em
{\large\it Graduate School of Mathematics \\
Nagoya University, Nagoya, 464-8602, Japan}
\vskip 0.8cm
{\large Yukinori Yasui\footnote{e-mail: yasui@sci.osaka-cu.ac.jp}}
\vskip 1.0em
{\large\it Department of Physics, Osaka City University \\
Sumiyoshi-ku, Osaka, 558-8585, Japan}
\end{center}
\vskip0.5cm

\begin{abstract}

Asymptotically locally conical (ALC) metric of exceptional holonomy
has an asymptotic circle bundle structure that accommodates
the $M$ theory circle in type IIA reduction.
Taking $Spin(7)$ metrics of cohomogeneity one as explicit examples,
we investigate deformations of ALC metrics, in particular
that change the asymptotic $S^1$ radius related to the type IIA string
coupling constant. When the canonical four form of $Spin(7)$ holonomy is
taken to be anti-self-dual, the deformations of $Spin(7)$ metric are
related to the harmonic self-dual four forms,
which are given by solutions to a system of first order differential
equations, due to the metric ansatz of cohomogeneity one.
We identify the $L^2$-normalizable solution that deforms
the asymptotic radius of the $M$ theory circle.


\end{abstract}
\end{titlepage}

\renewcommand{\thefootnote}{\arabic{footnote}}
\setcounter{footnote}{0}

\section{Introduction}
\setcounter{equation}{0}

Asymptotically locally conical (ALC) metrics of $G_2$ and $Spin(7)$
holonomy \cite{CGLP3}\cite{BGGG}\cite{CGLP5}\cite{GS}\cite{KY1}\cite{KY2}
are expected to play an important role
in $M$ theory compactification with minimal supersymmetry.
The ALC metric has a circle bundle structure and the radius of the $S^1$ fiber
stabilizes asymptotically. Hence ALC metrics with exceptional holonomy may
be regarded as higher dimensional generalizations
of ALF metrics in four dimensions, typical examples of which are the Taub-NUT
metric and the Atiyah-Hitchin metric. In $M$ theory compactification
the asymptotic circle of ALC metric is identified with $M$ theory circle
whose radius is related to the string coupling of type IIA reduction.
In general ALC metric has a modulus of asymptotic radius of the $S^1$
fiber other than the modulus of overall scaling of Ricci-flat metric.
By deforming the radius of the asymptotic circle with the overall scale
parameter being fixed, we can interpolate the strong coupling $M$ theory limit
and the weak coupling region of perturbative IIA string theory. Namely when the
radius goes to infinity, an asymptotic conical (AC) metric
arises as a limit of ALC metric and it describes
a purely gravitational background of $M$ theory.
The limit of the other side is what is called
the Gromov-Hausdorff limit and the limiting metric is a direct product of a
Ricci-flat metric of lower dimensions and the Euclidean metric.
Such a picture gives a beautiful $M$ theory unification \cite{CGLP7} of
the resolved conifold and the deformed conifold
of IIA background that has been employed to uncover the strong coupling
dynamics of $N=1$ supersymmetric gauge theory
in four dimensions \cite{AC1}\cite{AMV}\cite{AW}.
(See also \cite{AC2} for an early attempt.)
The AC metric in the strong coupling limit is a resolution of the cone metric
over $SU(2)^3/SU(2)$ that possesses a triality symmetry \cite{AW}.
Each of three metrics permutated by the triality has a different
Gromov-Haussdorff limit whose Calabi-Yau part is identified as
either the deformed conifold or one of the two resolved conifold metrics.
They are different resolutions of the cone metric over $SU(2)^2/U(1)$
related by flop operation.

In this article we will investigate the deformation of ALC metrics of
$Spin(7)$ holonomy with a view to $M$ theory compactification
described above. It is known that the metric of $Spin(7)$ holonomy is
characterized by a closed four form
\beq
\Omega = \Omega_{abcd} e^a \wedge e^b \wedge e^c \wedge e^d~, \label{canonical}
\eeq
where $\{e^a\}$ is a vielbein (an orthonormal frame) and $\Omega_{abcd}$ is
related to the structure constants of octonions \cite{Book}.
We will take a convention that the canonical four form $\Omega$ is {\it
anti-self-dual}. Then the formal dimensions of the moduli space of
metric of $Spin(7)$ holonomy is $1+ b_4^+$, where $b_4^+$ is the Betti
number of {\it self-dual} four form \cite{GPP}\cite{Book}.
Note that the overall scaling of Ricci-flat metric always gives
one dimensional modulus. The appearance of $b_4^+$ is understood as follows;
In eight dimensions the representation of the traceless part of metric tensor
is ${\bf 35}_V$ of $SO(8)$. When we have a metric of $Spin(7)$ holonomy,
there is a covariantly constant spinor in ${\bf 8}_C$ of $SO(8)$ and we have
a global identification of the vector representation and the spinor
representation ${\bf 8}_V \equiv {\bf 8}_S$.
This implies a further identification of ${\bf 35}_V$
with the representation ${\bf 35}_S$ of the self-dual four form.
On the other hand the representation ${\bf 35}_C$ of
the anti-self-dual four form is reducible and decomposed
as ${\bf 1}$ $\oplus$ {\bf 7} $\oplus$ {\bf 27}, because the conjugate 
spinor bundle is
reducible (${\bf 8}_C = {\bf 1} \oplus {\bf 7}$) on the manifold of
$Spin(7)$ holonomy. The seven dimensional invariant subspace
in ${\bf 35}_C$ is specified
by the canonical anti-self-dual four form $\Omega$ defined by 
(\ref{canonical}).
When a self-dual four form in ${\bf 35}_S$ is closed and hence harmonic,
the corresponding infinitesimal
deformation of the traceless part of metric in ${\bf 35}_V$
does not break the Ricci-flatness. Therefore, we can define
infinitesimal deformations of the traceless part of $Spin(7)$ metric
from harmonic self-dual four forms.
Precisely speaking this argument is valid for compact manifolds.
In this paper we consider $Spin(7)$ metrics of cohomogeneity one and
they are metrics on non-compact manifolds. Our claim of
the correspondence of harmonic self-dual four forms and deformations
of $Spin(7)$ metric remains formal. It is likely that we need
some conditions on the asymptotic behavior
of harmonic four forms, like $L^2$-normalizability.
Quite recently a general theory of $L^2$-harmonic forms on (non-compact)
gravitational instantons is developed in \cite{HHM}. It is an interesting
problem to explore the relation of $L^2$-normalizable harmonic forms
and deformations of non-compact Ricci-flat metrics
from mathematically more general view points.

This paper is organized as follows; throughout the paper we consider ALC
metrics of $Spin(7)$ holonomy that are of cohomogeneity one
with the principal orbit $Sp(2)/Sp(1)$ or $SU(3)/U(1)$.
In section two we describe invariant forms
on the principal orbit $G/K= Sp(2)/Sp(1)$ and $SU(3)/U(1)$
as the $K$-invariant subalgebra $\Lambda_{\rm inv}$ of the differential algebra
generated by the Maurer-Cartan forms of $G$. In the case of $SU(3)/U(1)$,
$\Lambda_{\rm inv}$ depends
on the choice of $U(1)$ subalgebra in the maximal torus
determined by a pair of integers $(k,\ell)$.
For the special embedding with $k=\ell$, $\Lambda_{\rm inv}$ becomes much
larger and we find that as a differential algebra it coincides with that of
$Sp(2)/Sp(1)$ model supplemented by a single non-trivial cohomology class.
Due to this relation the $SU(3)/U(1)$ model
with the special embedding of $U(1)$
is mostly parallel with $Sp(2)/Sp(1)$ model.
In section three we define an infinitesimal deformation of the
canonical $Spin(7)$ four form as a shift by the invariant self-dual
four forms. We will work out the induced change
in the orthonormal frame to make the deformed four form get back
into the canonical form.
The deformation is classified to two types. The first type deformation leads
only rescalings of the vielbein and thus maintains the diagonal metric ansatz.
On the other hand the second type implies the mixing of vielbein and
the corresponding metric gets more involved.
In section four we derive a system of first order differential equations
for the self-dual four form defining a deformation
to be closed\footnote{By (anti-)self-duality closed forms
are automatically harmonic.} in the case of background metric of
cohomogeneity one. According to the two types of deformation we call
the first order system $u$-system and $v$-system, respectively.
We present explicit solutions to the first order system
in section five. The $L^2$-normalizability of the solutions is examined.
Due to the qualitative difference of asymptotic behavior the condition
for $L^2$-normalizability in ALC metrics is weaker than the corresponding
AC metrics. We find that ALC metrics allow $L^2$-normalizable
self-dual four forms, but AC metrics do not in general.
By looking at power series expansion we identify the solution
that deforms the radius of asymptotic circle in ALC metric.
We conclude that the $L^2$ normalizable self-dual harmonic four form controls
the change in the radius of $M$-theory circle.
It would be worth mentioning that the $L^2$ normalizable harmonic four
forms of the opposite duality also play a role in the deformation of
supersymmetric background of $Spin(7)$ holonomy.
As is shown in \cite{ADG} an anti-self-dual harmonic four form
in {\bf 27} of the irreducible decomposition of ${\bf 35}_C$ provides
a flux for brane resolution and one can construct
a supersymmetric $M2$ brane solution \cite{CLP}\cite{CGLP2}.
In summary a self-dual harmonic form gives a purely gravitational deformation,
while an anti-self-dual harmonic form introduces a four form flux.

There are three appendixes at the end of paper.
Appendix A provides our convention of $SU(3)$ Maurer-Cartan equations.
In Appendix B we summarize a formalism of Hitchin on stable forms and
metrics of exceptional holonomy, which gave a background of our work.
Finally we consider the deformation of the Atiyah-Hitchin metric
in Appendix C, since it allows a description similar to
our approach in this paper. But there is a crucial difference.
The deformation of the Atiyah-Hitchin metric breaks the ansatz of
cohomogeneity one, while the deformation of ALC metrics in this
paper does not.

\section{Algebra of Invariant Forms}
\setcounter{equation}{0}

The geometry of $G_2$ and $Spin(7)$ structures is closely related to
three and four forms on a manifold $M$. Recently Hitchin has shown that
such a relation is described in terms of certain functionals on the space
of three and four forms \cite{Hitchin}\footnote{A brief summary of
his formalism is given in Appendix B.}. In the case of $G_2$ manifolds
his method has turned out to be a powerful tool (see \cite{GYZ} for example).
The space of differential forms on $M$ is usually infinite dimensional, 
but when the manifold $M$ admits an action of Lie group $G$ with the 
principal orbit $G/K$ of codimension $n$, one can make
an ansatz of the three and four forms based on the algebra of
invariant forms on $G/K$ and the problem can
be reduced to a finite dimensional one by homogeneity.
This approach is quite general and has been employed in \cite{Brand} for 
constructing the most general $G_2$ metrics with the principal
orbit $S^3 \times S^3$.
There are two homogeneous spaces $S^7 = Sp(2)/Sp(1)$
(the Hopf fibration of $S^7$) and $M_{k,\ell}:=SU(3)/U(1)_{k,\ell}$
(the Aloff-Wallach space) that are known to
allow a weak $G_2$ structure \cite{FKMS}. 
We can consider $Spin(7)$ metrics of cohomogeneity one 
where one of these homogeneous spaces is served as 
the principal orbit of codimension one.
In general the algebra of
invariant differential forms on a homogeneous space $G/K$ is
identified as the $K$-invariant subalgebra of the exterior algebra
(freely) generated by the Maurer-Cartan one forms of $G$.
Let us consider the algebra of $K$-invariant forms in each case separately.

\subsection{$Sp(2)/Sp(1)$}

As a $K=Sp(1)$ module the cotangent space of $S^7=Sp(2)/Sp(1)$ is decomposed as
\beq
T_e^* S^7= P_1 \oplus P_2 \oplus P_3 \oplus P_4~,
\eeq
where $P_i~(i=1,2,3)$ is one dimensional and $P_4$ is four dimensional.
Let $\sigma_i~(i=1,2,3)$ be a basis of $P_i$. They are $K$-invariant and
generate
another $Sp(1)$ subgroup of $Sp(2)$ that is commuting with $K=Sp(1)$.
We take a basis $\Sigma_\mu~(\mu=0,1,2,3)$ of $P_4$ such that
$ds^2 = \Sigma_0^2+\Sigma_1^2+\Sigma_2^2+\Sigma_3^2$ gives a standard
(conformal) metric on
the base $S^4$ of the $S^3$ Hopf fibration of $S^7$. The self-dual
combination of $\Sigma_\mu$ gives
$K$-invariant two forms;
\beqa
\omega_1 &=& \Sigma_0 \wedge \Sigma_1 + \Sigma_2 \wedge \Sigma_3~, \CR
\omega_2 &=& \Sigma_0 \wedge \Sigma_2 + \Sigma_3 \wedge \Sigma_1~,  \\
\omega_3 &=& \Sigma_0 \wedge \Sigma_3 + \Sigma_1 \wedge \Sigma_2~. \nonumber
\eeqa
We note the relations
\beqa
\omega_1 \wedge \omega_1 &=& \omega_2 \wedge \omega_2 = \omega_3 \wedge
\omega_3 =
2~\Sigma_0 \wedge \Sigma_1 \wedge \Sigma_2 \wedge \Sigma_3~, \CR
\omega_1 \wedge \omega_2 &=& \omega_2 \wedge \omega_3 = \omega_3 \wedge
\omega_1 = 0~. \label{rel1}
\eeqa
The algebra $\Lambda_{\rm inv}$ of $K$-invariant forms is generated by
$\sigma_i$
and $\omega_i$ with the relation (\ref{rel1}).
The exterior derivative keeps the subalgebra $\Lambda_{\rm inv}$ and on the
generators we have
\beqa
d\sigma_i &=& \epsilon_{ijk} \sigma_j \wedge \sigma_k + \omega_i~, \CR
d\omega_i &=& 2 \epsilon_{ijk} \sigma_j \wedge \omega_k~. \label{Sp}
\eeqa
By these relations we see that the space of closed three forms
in $\Lambda_{\rm inv}$ is three dimensional and
the most general exact four form is given by
\beqa
\Phi &=& x_1 d(\sigma_1 \wedge \omega_1) + x_2 d(\sigma_2 \wedge \omega_2)
+ x_3 d(\sigma_3 \wedge \omega_3) + 2x_4 d (\sigma_1 \wedge \sigma_2 \wedge
\sigma_3) \CR
& &~~+ y_1 d(\sigma_2\wedge \omega_3 + \sigma_3\wedge \omega_2) +
y_2 d(\sigma_3\wedge \omega_1 + \sigma_1\wedge \omega_3)
+ y_3 d(\sigma_1\wedge \omega_2 + \sigma_2\wedge \omega_1)~, \CR
&=& 2 (x_1 + x_2 + x_3) \Sigma_0 \wedge \Sigma_1 \wedge \Sigma_2 \wedge
\Sigma_3
+2(x_4 + x_1 -x_2 -x_3) \omega_1 \wedge \sigma_2 \wedge \sigma_3
\label{Spform} \\
& &~~+2(x_4 - x_1 +x_2 -x_3) \omega_2 \wedge \sigma_3 \wedge \sigma_1
+2(x_4 - x_1 -x_2 +x_3) \omega_3 \wedge \sigma_1 \wedge \sigma_2 \CR
& &~~~+ 4y_1 ( \omega_3 \wedge \sigma_3 \wedge \sigma_1 +\omega_2 \wedge
\sigma_1
\wedge \sigma_2)
+ 4y_2 ( \omega_1 \wedge \sigma_1 \wedge \sigma_2 +\omega_3 \wedge
\sigma_2 \wedge \sigma_3) \CR
& &~~~+ 4y_3 ( \omega_2 \wedge \sigma_2 \wedge \sigma_3 +\omega_1 \wedge
\sigma_3 \wedge \sigma_1)~. \nonumber
\eeqa

\subsection{$SU(3)/U(1)_{k,\ell}$ (generic case $k\neq\ell$)}

The subgroup $U(1)_{k,\ell}$ of the homogeneous space
$M_{k,\ell}:=SU(3)/U(1)_{k,\ell}$
is represented by ${\rm diag.}~(e^{ik\theta}, e^{i\ell\theta},
e^{im\theta})$ with $k+\ell+m=0$.
We use the following notation for $SU(3)$ left invariant one forms;
\beqa
E^1&=&\sigma_1~, \quad E^2=\sigma_2~, \quad
E^3=\Sigma_1~, \quad E^4=\Sigma_2~, \CR
E^5&=&\tau_1~, \quad E^6=\tau_2~, \quad
E^7=T_A~, \quad E^8=T_B~,
\eeqa
where $E^8$ corresponds to the subgroup $U(1)_{k,\ell}$.
The Maurer-Cartan equations in our convention
are summarized in Appendix A, where we note the relation;
\beqa
\alpha_A &=& k~, \quad \beta_A=\ell,~ \quad \gamma_A=m~, \CR
\alpha_B &=&\ell - m ~, \quad \beta_B=m - k,~ \quad \gamma_B=k - \ell~.
\eeqa
As a $U(1)_{k,\ell}$ module the cotangent space of $M_{k,\ell}$ is
decomposed as
\beq
T_e^* M_{k,\ell}= P_0 \oplus P_{\ell-m} \oplus P_{m-k} \oplus P_{k-\ell}~,
\eeq
where $P_q$ has the $U(1)$ weight $q$.
A natural basis for the above decomposition is
given by $E^7, E^1 \pm i E^2, E^3 \pm i E^4, E^5 \pm i E^6$ with charges
$0, \pm (\ell-m),
\pm (m-k), \pm (k-\ell)$, respectively.
The invariant differential forms of $M_{k,\ell}$ can be identified as
the subalgebra of the $U(1)$ charge free part generated by
\beqa
T_A &=& E^7~, \CR
\omega_k &=& \frac{i}{2} (E^{2k-1} + iE^{2k}) \wedge  (E^{2k-1} -
iE^{2k})~, \\
\Omega_1 + i\Omega_2 &=& (E^1 + iE^2) \wedge (E^3 + iE^4) \wedge (E^5 + iE^6)~,
\nonumber
\eeqa
with the relations
\beqa
(T_A)^2 &=& (\omega_i)^2 = (\Omega_j)^2 = 0~, \CR
\omega_i \wedge \Omega_j &=& 0~, \quad
\Omega_1 \wedge \Omega_2 = 4~\omega_1 \wedge \omega_2 \wedge \omega_3~.
\eeqa
The exterior derivative keeps the subalgebra and on the generators we have
\beqa
dT_A &=& 2k \omega_1 + 2\ell \omega_2 + 2m \omega_3~, \quad
d\omega_1 = d\omega_2 = d\omega_3 = \Omega_2~, \CR
d\Omega_1 &=& 4( \omega_1 \wedge \omega_2 + \omega_2 \wedge \omega_3 +
\omega_3 \wedge \omega_1)~, \quad
d\Omega_2 = 0~.
\eeqa
The space of closed three forms is one dimensional and
the most general exact four form is given by
\beqa
\Phi &=& x_1 d(\omega_1 \wedge T_A) + x_2 d(\omega_2 \wedge T_A)
+ x_3 d(\omega_3 \wedge T_A) + x_4 d\Omega_1~, \CR
&=& (4x_4 + 2\ell x_1 + 2k x_2) \omega_1 \wedge \omega_2
+ (4x_4 + 2m x_2 + 2\ell x_3) \omega_2 \wedge \omega_3 \\
& &~~~+(4x_4 + 2m x_1 + 2k x_3) \omega_3 \wedge \omega_1
- (x_1 + x_2 + x_3) T_A \wedge \Omega_2~. \nonumber
\eeqa

\subsection{$SU(3)/U(1)_{1,1}$ (special case $k=\ell$)}

When $k=\ell$, $E^5$ and $E^6$ become $U(1)$ singlets and $E^1 \pm i E^2$ and
$E^3 \pm i E^4$ have $U(1)$ charge $\pm 3$ and $\mp 3$, respectively.
The space of $K$-invariant forms gets much larger and is generated by $E^5,
E^6, E^7$ and
$\omega_1, \omega_2, \tilde\Omega_1, \tilde\Omega_2$. We have introduced
a $K$-invariant two form
\beq
\tilde\Omega_1 + i\tilde\Omega_2 = (E^1 +i E^2) \wedge (E^3 +i E^4)~,
\eeq
which satisfies $\tilde\Omega_1 \wedge \tilde\Omega_1 = \tilde\Omega_2
\wedge \tilde\Omega_2
= 2\omega_1 \wedge \omega_2$ and $\omega_i \wedge \tilde\Omega_j
= \tilde\Omega_1 \wedge \tilde\Omega_2 = \omega_1 \wedge \omega_1 =
\omega_2 \wedge \omega_2 =0$.
The exterior derivative on the generators is given by
\beqa
dE^5 &=& \tilde\Omega_1 - E^6 \wedge E^7, \quad
dE^6 = - \tilde\Omega_2 - E^7 \wedge E^5, \quad
dE^7 = 2\omega^+ -4 E^5 \wedge E^6, \CR
d\omega^+ &=& 2 \left(\tilde\Omega_2 \wedge E^5 +
\tilde\Omega_1 \wedge E^6\right)~, \quad
d\omega^- = 0~, \label{basis} \\
d\tilde\Omega_1 &=& -2\omega^+ \wedge E^6
- \tilde\Omega_2 \wedge E^7~, \quad
d\tilde\Omega_2 =  -2\omega^+ \wedge E^5
+ \tilde\Omega_1 \wedge E^7~, \nonumber
\eeqa
where $\omega^\pm := \omega_1 \pm \omega_2$.
In contrast with previous two cases there are non-trivial cohomology
class $[\omega^-]$ with degree two and its dual with degree five.
One can identify $[\omega^-]$ with the K\"ahler form of ${\bf CP}(2)$
when the coset $SU(3)/U(1)_{1,1}$ is viewed as the total space of a principal
$SO(3)$ bundle over ${\bf CP}(2)$.
It is amusing to compare the differential algebra (\ref{basis})
with that of $Sp(2)/Sp(1)$.
By the following mapping
\beqa
E^5 &\to& -\sigma_1~, \quad E^6 \to -\sigma_2~, \quad E^7 \to -2\sigma_3~, \CR
\tilde\Omega_1 &\to& -\omega_1~,  \quad \tilde\Omega_2 \to \omega_2~, \quad
\omega^+\to -\omega_3~, \label{mapping}
\eeqa
we find exactly the same differential algebra as (\ref{Sp}).
Hence at the level of the differential algebra of invariant forms,
$SU(3)/U(1)_{1,1}$ is obtained from $Sp(2)/Sp(1)$
by augmenting the non-trivial cohomology class $[\omega^-]$.
This observation has the following meaning; the complex projective
space ${\bf CP}(2)$ is both K\"ahler and quaternionic K\"ahler\footnote{In
four dimensions the quaternionic K\"ahler manifold is nothing but
the self-dual Einstein manifold.}. On the other hand the four dimensional
sphere $S^4$ is quaternionic K\"ahler but non-K\"ahler. The triplets
$\omega_i~(i=1,2,3)$ for $S^4$ and $(\tilde\Omega_1, \tilde\Omega_2,
\omega^+)$ for ${\bf CP}(2)$ define a quaternionic K\"ahler structure
of each manifold and as remarked above $\omega^-$ defines a K\"ahler
structure of ${\bf CP}(2)$. Note that the K\"ahler two form $\omega^-$
has the opposite duality to the triplet of two forms for the quaternionic
K\"ahler structue. This is explained from the holonomy
$U(2) = U(1)_L \times SU(2)_R$ of ${\bf CP}(2)$. Namely
the K\"ahler two form is the curvature of the line bundle
associated with $U(1)_L$, while it is the curvature of
$SO(3)_R = SU(2)_R/ {\bf Z}_2$ bundle which gives
the quaternionic K\"ahler two forms.
(Since ${\bf CP}(2)$ is not spin, there is
no lift of the $SO(3)_R$ bundle to $SU(2)_R$ bundle.)

We find the space of closed three forms is three dimensional. A basis
is given by exact three forms $d\omega^+, d\tilde\Omega_1$ and $d\tilde
\Omega_2$ in (\ref{basis}). Hence,
the most general exact four form has ten parameters.
Taking the above correspondence to $Sp(2)/Sp(1)$ model into account,
let us expand an invariant three form $\phi$ in the following form;
\beqa
\phi &=& x_1 (\tilde\Omega_1 \wedge E^5) - x_2 (\tilde\Omega_2 \wedge E^6)
+ \frac{1}{2} x_3 (\omega^+ \wedge E^7) - x_4 (E^5 \wedge E^6 \wedge E^7)~, \CR
& &~+ y_1 (\omega^+ \wedge E^5 -\frac{1}{2} \tilde\Omega_2 \wedge E^7)
+ y_2 (\omega^+ \wedge E^6 +\frac{1}{2}  \tilde\Omega_1 \wedge E^7)
+ y_3 ( - \tilde\Omega_2 \wedge E^5 + \tilde\Omega_1 \wedge E^6 ) \CR
& &~+ z_1 (\omega^- \wedge E^5) + z_2 (\omega^- \wedge E^6)
+ \frac{1}{2} z_3 (\omega^- \wedge E^7)~.
\eeqa
Then the corresponding exact four form is
\beqa
\Phi &:=& d \phi \CR
&=& (x_1+x_2+x_3) \omega^+ \wedge \omega^+
- (x_4+x_1-x_2-x_3) \tilde\Omega_1 \wedge E^6 \wedge E^7 \CR
& &~+ (x_4-x_1+x_2-x_3) \tilde\Omega_2 \wedge E^7 \wedge E^5
-2 (x_4-x_1-x_2+x_3) \omega^+ \wedge E^5 \wedge E^6 \CR
& &~+ 2y_1 \left(- \omega^+ \wedge E^7 \wedge E^5
+2 \tilde\Omega_2 \wedge E^5 \wedge E^6 \right)
+ 2y_2 \left( -2\tilde\Omega_1 \wedge E^5 \wedge E^6
- \omega^+ \wedge E^6 \wedge E^7 \right) \CR
& &~+2y_3 \left( \tilde\Omega_2 \wedge E^6 \wedge E^7
- \tilde\Omega_1 \wedge E^7 \wedge E^5 \right) \CR
& &~-z_1(\omega^- \wedge E^6 \wedge E^7) -z_2 (\omega^- \wedge E^7 \wedge E^5)
-2z_3 (\omega^- \wedge E^5 \wedge E^6)~. \label{SU3basis}
\eeqa

\section{Deformation of Four Form and  ALC metric}
\setcounter{equation}{0}

A metric with special holonomy is Ricci flat and an overall scaling of the
metric always gives one dimensional trivial modulus of Ricci flat metrics.
When a Ricci-flat metric of cohomogeneity one is
obtained as a resolution of the cone
metric over $G/K$, this scale parameter controls the resolution of conical
singularity and it is related to the volume of the bolt singularity;
$S^4$ for $Sp(2)/Sp(1)$ case and ${\bf CP}^2$ for $SU(3)/U(1)$ case.
Hence we can fix the modulus of scaling by normalizing
the size of the bolt singularity.
Then non-trivial moduli come from deformations of the traceless part of metric.
When the manifold is compact, infinitesimal deformations that keep
Ricci-flatness are
given by zero modes of the Lichnerowicz laplacian \cite{GPP}.
A reduction of the holonomy to special holonomy allows us to identify these
zero modes with zero-modes of exterior derivative on differential
form (harmonic forms) of an appropriate degree.
For example for $G_2$ holonomy the formal dimensions of $G_2$ metrics is
given by the third Betti number $b_3$ and for $Spin(7)$
the dimensions is given by $1+ b_4^+$ where $b_4^+$ is the Betti number
of self-dual four forms\footnote{Our convention is
that the canonical four form of $Spin(7)$ structure is a closed anti-self-dual
four form. If it is self-dual, $b_4^+$ should be replaced by $b_4^-$; the
Betti number of anti-self-dual four forms.}\cite{Book}.
For non-compact manifold the correspondence of
infinitesimal deformations of Ricci flat metrics and
the harmonic forms becomes subtle.
We may impose some conditions on normalizablity and/or
asymptotic behavior on differential
forms. The condition we should impose is not necessarily unique and
it would characterize
the physical meaning of corresponding moduli. For example the moduli
of metrics that come from $L^2$-normalizable closed forms will not
change the asymptotic behavior and are expected to be dynamical,
while those from $L^2$-non-normalizable forms are non-dynamical.
They can change the asymptotic behavior of the metric and
are regarded as a change of background metric.
In the following based on the coset geometry of $Sp(2)/Sp(1)$ and $SU(3)/U(1)$,
we will see quite explicitly the correspondence between the harmonic
self-dual four forms and infinitesimal deformations of $Spin(7)$ metric.
In our models of $Sp(2)/Sp(1)$ and $SU(3)/U(1)$ the fourth de Rham
cohomology of the coset is trivial and we will take exact four forms on the
coset to produce self-dual four forms on the total space.
Thus the harmonic $L^2$ normalizable four forms $G$ in this paper
is exact $G = d C$ in algebraic sense. However, it is non-trivial
in the sense of $L^2$-cohomology, since the three form potential $C$ is
not $L^2$ normalizable in general.

\subsection{$Sp(2)/Sp(1)$}

Assume that the vielbeins are given by
\beqa
e^1 &=& a(t)\Sigma_1~, \quad e^3 = a(t)\Sigma_2~, \quad e^5 = a(t)\Sigma_3~,
\quad e^7 = a(t) \Sigma_0~, \CR
e^2 &=& b_1(t)\sigma_1~, \quad e^4 = b_2(t)\sigma_2~, \quad
e^6 = b_3(t)\sigma_3~.
\eeqa
As the canonical ASD four form of $Spin(7)$ structure on
${\bf R}\times Sp(2)/Sp(1)$, we take
\beq
\Omega_0^{\rm ASD} = dt \wedge * \rho_0 - \rho_0~,
\eeq
where
\beqa
\rho_0 &=& - e^{1357} + e^{1467} + e^{3456} + e^{2367} + e^{1256}
+ e^{2457} + e^{1234}~, \CR
&=& \frac{1}{2} (e^{12} + e^{34} + e^{56})^2
- {\rm Re}\left[(e^1 + ie^2)(e^3 +ie^4)(e^5 +ie^6)\right] \wedge e^7~,
\label{4form} \\
&=& a^4 \Sigma_0 \wedge \Sigma_1 \wedge \Sigma_2 \wedge \Sigma_3
- a^2 b_2 b_3 (\omega_1 \wedge \sigma_2 \wedge \sigma_3)
- a^2 b_3 b_1 (\omega_2 \wedge \sigma_3 \wedge \sigma_1) \CR
& &~~- a^2 b_1 b_2 (\omega_3 \wedge \sigma_1 \wedge \sigma_2)~, \nonumber
\eeqa
and
\beqa
*\rho_0 &=& - e^{246} + e^{352} + e^{712} + e^{514} + e^{734} + e^{136} +
e^{756}~, \CR
&=& (e^{12} + e^{34} + e^{56}) \wedge e^7 + {\rm Im}\left[(e^1 + ie^2)(e^3
+ie^4)(e^5 +ie^6)\right]~, \label{3form} \\
&=& -b_1b_2b_3 (\sigma_1 \wedge \sigma_2 \wedge \sigma_3) + a^2 b_1
(\omega_1 \wedge \sigma_1)
+ a^2 b_2 (\omega_2 \wedge \sigma_2) + a^2 b_3 (\omega_3 \wedge \sigma_3)~.
\nonumber
\eeqa
The Hodge operator $*$ is defined by the coset metric $\hat{g}(t) = e^a
\otimes e^a$
and we have $\alpha\wedge *\alpha = e^{1234567}$ if $\alpha$ is
an exterior product of $e^a$.
When the four form $\Omega_0^{\rm ASD}$ is closed,
the $Spin(7)$ structure is called
torsion free and the corresponding metric $g=dt^2+\hat{g}(t)$
has $Spin(7)$ holonomy.
It is easy to see that
the octonionic instanton equation for the spin connection implies
$d\Omega_0^{\rm ASD}=0$.

Now let us look at an infinitesimal shift of the canonical
four form by a self-dual four form
\beqa
\Omega &=& \Omega_0^{\rm ASD} + \epsilon G^{\rm SD} \CR
&=& (dt \wedge *\rho_0 - \rho_0) + \epsilon( dt \wedge \Psi + \Phi) \\
&=& dt \wedge \sigma - \rho~, \nonumber
\eeqa
where $\sigma \equiv *\rho_0 + \epsilon \Psi$ is a three form and $\rho \equiv
\rho_0 - \epsilon \Phi$ is a four form.  The seven
dimensional duality $*\Phi=\Psi$  implies the eight dimensional
self-duality of $G^{\rm SD}=dt \wedge \Psi+\Phi$.
As before for the deformed four form
$\Omega$ to define a torsion free
$Spin(7)$ structure we must require that $d\Omega=0$.
Let us take the invariant four form $\Phi$ that is closed in seven
dimensional sense.
Hence in the $Sp(2)/Sp(1)$ model we take (see Eq.(\ref{Spform}));
\beqa
\Phi &=& u_1 a^4(\Sigma_0 \wedge \Sigma_1 \wedge \Sigma_2 \wedge \Sigma_3)
+ u_2 a^2 b_2 b_3 (\omega_1 \wedge \sigma_2 \wedge \sigma_3) \CR
& &~~+ u_3 a^2 b_3 b_1 (\omega_2 \wedge \sigma_3 \wedge \sigma_1)
+ u_4 a^2 b_1 b_2 (\omega_3 \wedge \sigma_1 \wedge \sigma_2) \CR
& &~~~ + v_1 ( \omega_2 \wedge \sigma_2 \wedge \sigma_3 +\omega_1 \wedge
\sigma_3 \wedge \sigma_1)
+ v_2 ( \omega_3 \wedge \sigma_3 \wedge \sigma_1 +\omega_2 \wedge \sigma_1
\wedge \sigma_2) \CR
& &~~~~~~+ v_3 ( \omega_1 \wedge \sigma_1 \wedge \sigma_2 +\omega_3 \wedge
\sigma_2 \wedge \sigma_3)
\eeqa
In our ansatz of cohomogeneity one metric the coefficients $u_i$ and $v_i$ are
functions of $t$ and $d\Phi = dt \wedge (d\Phi/dt)$ in eight dimensions.
$dG^{\rm SD}=0$ means $d\Phi/dt=d*\Phi$.
This condition gives a first order system for the coefficients $u_i$ and
$v_i$. We will describe it more explicitly in the next section.

When $v_i=0$ the diagonal metric ansatz is maintained. In fact assuming
that new vielbeins are
\beq
\tilde e^i = \left( 1 + \epsilon U_i \right) e^i~, \quad U_1=U_3=U_5=U_7
\eeq
we obtain in the leading order
\beqa
U_1 &=& \frac{1}{4} u_1~, \quad
U_2 = \frac{1}{2} ( -u_2 + u_3 + u_4) -\frac{1}{4} u_1~, \CR
U_4 &=& \frac{1}{2} ( u_2 - u_3 + u_4) -\frac{1}{4} u_1~, \quad
U_6 = \frac{1}{2} ( u_2 + u_3 - u_4) -\frac{1}{4} u_1~.
\eeqa
together with a redefinition of \lq\lq time\rq\rq\ variable
\beq
d\tilde t = \left( 1 - \epsilon (4U_1 + U_2+U_4+U_6) \right) dt~.
\eeq
We can see that $\Omega$ is transformed into the canonical form in terms of
these new vielbeins.

On the other hand non-vanishing $v_i$ causes a \lq\lq mixing\rq\rq\ of the
original vielbein. For example one can consider the deformation by
\beqa
\Phi_{mix} &=& v_1 \left( b_1(e^{7346} + e^{5146}) + b_2(e^{7162} +
e^{3562}) \right) \CR
& &+v_2 \left( b_2(e^{7562} + e^{1362}) + b_3(e^{7324} + e^{5124}) \right) \CR
& &+v_3 \left( b_3(e^{7124} + e^{3524}) + b_1(e^{7546} + e^{1346}) \right)~,
\eeqa
where we have scaled the infinitesimal deformation parameters $v_i$.
We have checked that the deformed four form can be transformed into
the canonical form by allowing the \lq\lq mixing\rq\rq\ of the orthonormal
frame. Namely we find a solution to the condition
\beq
\Omega_0^{ASD} + \epsilon (\Phi_{mix} + dt\wedge *\Phi_{mix})
= dt\wedge \tilde{*} \tilde\rho_0 - \tilde\rho_0
\eeq
where $\tilde \rho_0$ and $\tilde{*} \tilde\rho_0$ are given by (\ref{4form})
and (\ref{3form}) with $e^i$ being replaced
by a new orthonormal frame $\tilde e^i$.
The new orthonormal frame $\tilde e^i$ is assumed to be a linear
combination of the original vielbein.
Furthermore one is forced to change the frame of the radial
direction $e^0 = dt$ mixed with the frames $e^i$ of the principal orbit.
Due to this mixing of the radial geodesic direction and the principal orbit,
it is not clear if the deformed metric remains of cohomogeneity one.
As we will see later in explicit examples there are no $L^2$-normalizable
solution providing
the deformations of this type (see the $v$-system in the next section)
and we will leave this issue as an open problem.

\subsection{$SU(3)/U(1)_{1,1}$ (special case $k=\ell$)}

In this case we take the following vielbeins given by
\beqa
e^1 &=& a(t) E^1~, \quad e^2 = a(t) E^2~, \quad e^3 = b(t) E^3~,
\quad e^4 = b(t) E^4~, \CR
e^5 &=& c_{1}(t)E^5~, \quad e^6 = c_{2}(t) E^6~, \quad e^7 = f(t) E^7~.
\label{SU3frame}
\eeqa
and the canonical ASD four form of $Spin(7)$ structure on
${\bf R}\times SU(3)/U(1)_{1,1}$
\beq
\Omega_0^{\rm ASD} = dt \wedge * \rho_0 - \rho_0~,
\eeq
where
\beqa
\rho_0 &=&  e^{1234} + e^{1256} + e^{3456} - e^{1367} + e^{2467}
- e^{1457} - e^{2357}~, \CR
&=& \frac{1}{2} (e^{12} + e^{34} + e^{56})^2
- {\rm Im}\left[(e^1 + ie^2)(e^3 +ie^4)(e^5 +ie^6)\right] \wedge e^7~, \\
&=& \frac{1}{2} a^2 b^2 (\omega^+ \wedge \omega^+)
+ \frac{1}{2} (a^2+b^2) c_1c_2 (\omega^+ \wedge E^5 \wedge E^6)
+ \frac{1}{2} (a^2-b^2) c_1c_2 (\omega^- \wedge E^5 \wedge E^6) \CR
& &~~+ abc_2 f (-\tilde\Omega_1 \wedge E^6 \wedge E^7) +
abc_1 f (\tilde\Omega_2 \wedge E^7 \wedge E^5)~, \nonumber
\eeqa
and
\beqa
*\rho_0 &=&  e^{567} + e^{347} + e^{127} + e^{315} + e^{524} + e^{461} +
e^{362}~, \CR
&=& (e^{12} + e^{34} + e^{56}) \wedge e^7 - {\rm Re}\left[(e^1 + ie^2)(e^3
+ie^4)(e^5 +ie^6)\right]~, \\
&=& c_1 c_2 f (E^5 \wedge E^6 \wedge E^7)
+ \frac{1}{2} (a^2+b^2) f (\omega_+ \wedge E^7)
+ \frac{1}{2} (a^2-b^2) f (\omega_- \wedge E^7) \CR
& & ~~+ ab c_1 (-\tilde\Omega_1 \wedge E^5) + ab c_2(\tilde\Omega_2 \wedge
E^6)~.
\nonumber
\eeqa
An infinitesimal deformation of $Spin(7)$ four form is defined by
considering the
following most general ansatz of the invariant four form $\Phi$;
\beqa
\Phi &=& u_{1}(t)e^{1234}+u_{2}(t)e^{1256} +u_{3}(t)e^{3456}
+ u_{4}(t)(-e^{1367}+e^{2467})+ u_{5}(t)(-e^{2357}-e^{1457}) \CR
&+& v_{1}(t)e^{1267}-v_{2}(t)e^{1257}+
v_{3}(t)e^{3467} - v_{4}(t)e^{3457}+v_{5}(t)(e^{2457}-e^{1357}) \\
&+& \ell_{1}(t)(-e^{2456}+e^{1356})+ \ell_{2}(t)(e^{2356}+e^{1456})
+ \ell_{3}(t)(e^{2367}+e^{1467}). \nonumber
\eeqa
We can check that if we impose the condition
\beq
\ell_1 = \frac{f}{a b c_1}(a^2 v_1+b^2 v_3)~, \quad
\ell_2 = -\frac{f}{a b c_2}(a^2 v_2+b^2 v_4)~, \quad
\ell_3 = -\frac{c_1}{c_2}v_5,
\eeq
then $\Phi$ is closed form on $SU(3)/U(1)$ and expressed as follows (see 
Eq.(\ref{SU3basis}));
\beqa
\Phi &=& \frac{1}{2} u_1 a^2 b^2 (\omega^+ \wedge \omega^+)
- u_4 abc_2f (\tilde\Omega_1 \wedge E^6 \wedge E^7)
+ u_5 abc_1f (\tilde\Omega_2 \wedge E^7 \wedge E^5) \CR
&+& \frac{1}{2} (u_2 a^2 + u_3 b^2) c_1c_2 (\omega^+ \wedge E^5 \wedge E^6)
+ abc_1 f v_5 (\tilde\Omega_1 \wedge E^7 \wedge E^5
- \tilde\Omega_2 \wedge E^5 \wedge E^6) \CR
&+& c_1f (v_2 a^2 + v_4 b^2) \left( \frac{1}{2} \omega^+ \wedge E^7 \wedge E^5
- \tilde\Omega_2 \wedge E^5 \wedge E^6 \right)  \CR
&+& c_2f (v_1 a^2 + v_3 b^2) \left( \frac{1}{2} \omega^+ \wedge E^6 \wedge E^7
+ \tilde\Omega_1 \wedge E^5 \wedge E^6 \right) \label{SU3Phi} \\
&+& \frac{1}{2} (u_2 a^2 - u_3 b^2) c_1c_2 \omega^- \wedge E^5 \wedge E^6
+ \frac{1}{2} (v_1 a^2 - v_3 b^2) c_2 f \omega^- \wedge E^6 \wedge E^7 \CR
&+& \frac{1}{2} (v_2 a^2 - v_4 b^2) c_1 f \omega^- \wedge E^7 \wedge E^5~.
\nonumber
\eeqa
We find the parameters $u_i$ give rise to a diagonal deformation of
the metric $\tilde g =g+\epsilon \, h_{diag}/2$~($\epsilon \ll 1$).
By using the orthonormal basis $g=dt^2+e^a \otimes e^a$,
the metric is written as
\beq
h_{diag}=h_0 \,dt^2+\sum_{a=1}^{7} h_a e^a \otimes e^a~,
\eeq
where
\beqa
h_0 &=& u_1 +u_2 + u_3 +2(u_4 + u_5)~, \quad
h_1 = h_2 =-u_1 - u_2 + u_3~, \CR
h_3 &=& h_4 =-u_1 + u_2 - u_3~, \quad
h_5 = u_1 - u_2 - u_3 +2(u_4 - u_5)~, \CR
h_6 &=& u_1 - u_2 - u_3 -2(u_4 - u_5)~, \quad
h_7 = u_1 + u_2 + u_3 -2(u_4 + u_5)~.
\eeqa
We note that by the correspondence $u_1 \to u_1, u_2=u_3 \to u_4, u_4 \to
u_2, u_5 \to u_3$ the result of $Sp$ model is reproduced up to an overall
factor.

\section{First Order System for Harmonic  Four Forms}
\setcounter{equation} {0}

In this section we will give the condition for the self-dual four form
$G^{SD}$ to be closed and hence to be harmonic. 
When an eight-manifold $(M, g)$ is of cohomogeneity one,
that is, $M$ admits an action of the Lie group $G$ 
with seven dimensional principal orbits $G/K$, 
the manifold is locally $M \simeq {\bf R} \times G/K$.
Taking a unit vector field normal to the orbit, 
we can write the metric in the form
\beq
g = dt^2+\hat{g}(t)~,
\eeq
where $\hat{g}(t)$ is a $G$-invariant metric on the orbit $G/K$.
With this metric the condition for $G^{SD}$ is expressed by a system of first order
differential equations for the coefficients of invariant form on $G/K$.

\subsection{$SU(3)/U(1)_{1,1}$}

Our choice of the vielbeins (\ref{SU3frame}) implies the following diagonal
form
of $\hat{g}(t)$ for all $t$~;
\beqa
\hat{g}(t) &=& a(t)^2(E^1 \otimes E^1+E^2 \otimes E^2)+b(t)^2(E^3 \otimes E^3
+E^4 \otimes E^4) \CR
& & +c_1(t)^2 E^5 \otimes E^5 +c_2(t)^2 E^6 \otimes E^6
+f(t)^2 E^7 \otimes E^7. \label{SU3metric}
\eeqa
Let us consider a self-dual 4-form on $M$ of the following form;
\beq
G^{SD} = \Phi+ e^0  \wedge * \Phi~, \quad (e^0:= dt)~,
\label{ansatz}
\eeq
where $\Phi$ is expanded by a basis for exact invariant four-forms on $G/K$
and an explicit form is given by (\ref{SU3Phi}).
Since $\Phi$ in (\ref{ansatz}) is exact on $G/K$, the closeness
condition $dG^{SD}=0$ is expressed by
\beq
\frac{d}{dt}\Phi = d * \Phi~. \label {grad}
\eeq
   From the result in section 2.3 we see that $* \Phi$ is a $K$-invariant
three form and $d * \Phi$ can be expanded by the basis for
exact four forms given in section 2.3.
Thus, (\ref{grad}) yields the following
first order differential equations;

(A) $u$-system
\beqa
\frac{d}{dt}&(a^2& b^2 u_1)
-2a^2 f u_3-2b^2fu_2+2abc_1 u_4+2abc_2 u_5=0~, \CR
\frac{d}{dt}&(a^2& c_1 c_2 u_2)
+4a^2 f u_3-2c_1 c_2 fu_1+2abc_1 u_4+2abc_2 u_5=0~, \CR
\frac{d}{dt}&(b^2& c_1 c_2 u_3)
+4b^2 f u_2-2c_1 c_2 fu_1+2abc_1 u_4+2abc_2 u_5=0~, \label{form1} \\
\frac{d}{dt}&(ab& c_2 fu_4+abc_1 fu_5)+2f(a^2 u_3+b^2 u_2+c_1 c_2 u_1)=0~, \CR
\frac{d}{dt}&(ab& c_2 fu_4-abc_1 fu_5)+
2(abc_1 u_4-abc_2 u_5)=0~. \nonumber
\eeqa

(B) $v$-system
\beqa
\frac{d}{dt}&(a^2& c_2 f v_1)+(a^2 c_1+2b^2 f^2/c_1)v_3
+(2a^2 f^2/c_1)v_1=0~, \CR
\frac{d}{dt}&(b^2& c_2 f v_3)+(b^2 c_1+2a^2 f^2/c_1)v_1
+(2b^2 f^2/c_1)v_3=0~, \CR
\frac{d}{dt}&(a^2& c_1 f v_2)+(a^2 c_2+2b^2 f^2/c_2)v_4
+(2a^2 f^2/c_2)v_2=0~, \label{form2} \\
\frac{d}{dt}&(b^2& c_1 f v_4)+(b^2 c_2+2a^2 f^2/c_2)v_2
+(2 b^2 f^2/c_2)v_4=0~, \CR
\frac{d}{dt}&(ab& c_1 f v_5)+\frac{ab}{c_2}(c_1^2+c_2^2)v_5=0.
\nonumber
\eeqa
We note that the first order system for $u_i$  decouples from that for $v_i$.
Furthermore the $v$-system decomposes into three independent systems
for $(v_1,v_3),~(v_2,v_4)$ and $v_5$. The $(v_2,v_4)$ system is obtained
from the $(v_1,v_3)$ system by the replacement $c_2 \to c_1$.

\subsection{$SU(3)/U(1)_{k,\ell}$ ($k\neq\ell$)}

The $u$-system in the generic case $(k \neq \ell)$ can be obtained
similarly. The invariant metric $\hat{g}$ on the coset
is given by (\ref{SU3metric}) with $c \equiv c_1 = c_2$ and
according to the result in section 2.2, we put
$u_4 = u_5$, $v_i = \ell_j =0$ in the expansion of $G^{SD}$ in terms
of invariant forms. Then the $u$-system is reduced to
\beqa
\frac{d}{dt}(a^2 b^2 u_1) &=& -4abcu_4+2kfb^2 u_2+2\ell fa^2 u_3~, \CR
\frac{d}{dt}(a^2 c^2 u_2) &=& -4abcu_4+2kfc^2 u_1+2mfa^2 u_3~, \CR
\frac{d}{dt}(b^2 c^2 u_3) &=& -4abcu_4+2\ell fc^2 u_1+2mfb^2 u_2~,
\label{usys} \\
\frac{d}{dt}(abcf u_4) &=& -f(a^2 u_3 +b^2 u_2 +c^2 u_1)~. \nonumber
\eeqa
This equation has been derived in \cite{KY1}, but the overall sign of
the right-hand side is reversed here. The sign is determined by
the duality of closed four forms. In \cite{KY1} we were interested in
the anti-self-dual closed four forms (in our present
convention\footnote{Unfortunately the convention of \cite{KY1}, which is
the same as \cite{KB}, is opposite to the present paper and \cite{CLP}.}) 
in order to
construct supersymmetric brane solutions
following \cite{KB}\cite{CLP}\cite{CGLP3}.
When the canonical $Spin(7)$ form $\Omega$ is taken to be
anti-self-dual, it is self-dual harmonic four forms that is relevant to
metric deformations. On the other hand anti-self-dual harmonic four
forms are used to construct brane solutions that preserve supersymmetry.
We also note that the equation for the reversed duality satisfied the
linear relation
$u_1+u_2+u_3+4u_4= {\rm const}$,
while there is no such a relation in (\ref{usys}).

\subsection{$Sp(2)/Sp(1)$}

Using the correspondence (\ref{mapping}) found in section two,
we can derive the first order system for $Sp(2)/Sp(1)$ model
rather easily. Recall that the metric ansatz is
\beq
g = dt^2 + a(t)^2 (\Sigma_0^2 +\Sigma_1^2 +\Sigma_2^2 +\Sigma_3^2)
+ b_1(t)^2 \sigma_1^2 + b_2(t)^2 \sigma_2^2 + b_3(t)^2 \sigma_3^2~.
\label{Spmetric}
\eeq
Let us begin with the $u$-system.
By substituting $a=b \to a, c_1 \to b_1 c_2 \to b_2, -2f \to b_3$
and making the replacement $u_1 \to u_1, u_2=u_3 \to u_4,
u_4 \to u_2, u_5 \to u_3$ we obtain

(A) $u$-system
\beqa
\frac{d}{dt}&(a^4& u_1)
+2a^2 (b_1 u_2+ b_2 u_3+ b_3 u_4)=0~, \CR
\frac{d}{dt}&(a^2& b_2 b_3 u_2)
+ b_1 b_2 b_3 u_1 +2 a^2 (- b_1 u_2 + b_2 u_3 + b_3 u_4) =0~, \CR
\frac{d}{dt}&(a^2& b_3 b_1 u_3)
+ b_1 b_2 b_3 u_1 +2 a^2 ( b_1 u_2 - b_2 u_3 + b_3 u_4) =0~,
\label{Spusystem} \\
\frac{d}{dt}&(a^2& b_1 b_2 u_4)
+ b_1 b_2 b_3 u_1 +2 a^2 ( b_1 u_2 + b_2 u_3 - b_3 u_4) =0~.
\nonumber
\eeqa
The $u$-system for $Sp(2)/Sp(1)$ model
becomes more symmetric in the sense that
it has a cyclic symmetry in $(u_2, u_3, u_4)$.

In the same way we can obtain the following $\tilde v$-system for $Sp(2)/Sp(1)$

(B) $\tilde v$-system
\beqa
\frac{d}{dt}&(a^2& b_2 b_3  \tilde v_1)
- \frac{2a^2}{b_1} (b_3^2 + b_1^2 ) \tilde v_1 =0~, \CR
\frac{d}{dt}&(a^2& b_3 b_1  \tilde v_2)
- \frac{2a^2}{b_2} (b_1^2 + b_2^2 ) \tilde v_2 =0~, \\
\frac{d}{dt}&(a^2& b_1 b_2  \tilde v_3)
- \frac{2a^2}{b_3} (b_2^2 + b_3^2 ) \tilde v_3 =0~, \label{vsys}
\nonumber
\eeqa
where we have made the identification $\tilde v_1 \equiv v_1=v_3$,
$\tilde v_2 \equiv v_5$ and $\tilde v_3 \equiv (b_2/b_3) v_2 = (b_2/b_3) v_4$.
Note that the $\tilde v$ system decouples completely into
three independent ODE and again we find a cyclic symmetry in $(\tilde v_1,
\tilde v_2, \tilde v_3)$.

\section{Examples}
\setcounter{equation}{0}

To solve the $u$-system and the $v$-system obtained in the last section 
we need an expression of the background metric $a(t), b(t), c_i(t)$ and $f(t)$.
Let us consider a few examples of explicit ALC metrics.
We also discuss AC metrics briefly, since qualitative features are different.

\begin{flushleft}
{\bf 1. ALC metric}
\end{flushleft}

The first example is a deformation of the ALC $Spin(7)$ metric
${\bf B}_{8}$ on the bundle of chiral spinors over $S^4$.
In this example the existence of deformation was established globally in
\cite{CGLP3},
as a family of complete metrics ${\bf B}_{8}^{\pm }$ is known explicitly.
Our calculation shows that the infinitesimal variation of the metric is
controlled
by an $L^2$-normalizable harmonic self-dual four form $G^{SD}$.
The second example is a deformation of the ALC $Spin(7)$ metric on the
$Spin^c$ bundle over ${\bf CP}(2)$ \cite{CGLP5}\cite{KY1}\cite{GS}.
In this case a non-trivial
deformation was suggested by numerical analyses \cite{CGLP5}\cite{KY1} and
the existence of an $L^2$-normalizable $G^{SD}$ gives a further evidence for
the deformation.

Let us start with the ${\bf B}_{8}$ metric. The metric is given by
(\ref{Spmetric}) with
\beq
a^2=r^2-\ell^2~, \quad b_{1}^2=b_{2}^2=(r-3\ell)(r+\ell)~, \quad
b_{3}^2=4\ell^2 (r-3\ell)(r+\ell)/(r-\ell)^2
\eeq
and $dt=(r-\ell)/\sqrt{(r-3\ell)(r+\ell)}~dr$~.
The radial coordinate $r$ runs from the singular orbit $S^4$ at $r=3\ell$
to infinity. From the $u$-system (\ref{Spusystem}) with $b_1=b_2$, it is
easy to find
\beq
u_2 -u_3=k_0 \frac{(r-3\ell)^3}{(r+\ell)^6} \exp(2r/\ell)~, \label{k0}
\eeq
with a constant $k_0$. Since this solution has exponential growth at infinity,
we will take $k_0=0$. A convenient way to obtain the remaining
solutions is to introduce a new variable $f=a^2 b_1 b_2 (u_2+u_3)$.
Taking derivatives of the first-order equations (\ref{Spusystem}), 
we obtain the following differential equation of Fuchsian type
after rescaling the radial coordinate $x=r/\ell$~;
\beq
\frac{d^{3}f}{dx^3}+p_1(x)\frac{d^{2}f}{dx^2}+p_2(x)\frac{df}{dx}+p_3(x)f=0~,
\eeq
where
\beqa
p_1 &=& \frac{2(3x^2-10x-1)}{(x+1)(x-1)(x-3)}~, \CR
p_2 &=& -\frac{6x^3-18x^2+58x-110}{(x+1)^2(x-1)(x-3)^2}~, \CR
p_3 &=& -\frac{8(3x^4-24x^3+46x^2-32x-9)}{(x+1)^3(x-1)(x-3)^3}~.
\eeqa
This equation can be integrated and we find three linearly independent
solutions
\beqa
f_1 &=& \frac{64(x-3)^3}{(x+1)^2(x-1)^2}~, \quad
f_2=\frac{3x-5}{(x+1)^2(x-1)^2(x-3)}~, \CR
f_3 &=& \frac{(x-3)^3(3x^4+24x^3+98x^2+288x+723)}{2(x+1)^2(x-1)^2}~.
\eeqa
The second solution $f_2$ is singular at the point $x=3$ which corresponds
to the singular orbit $S^4$ and hence the regular solution in the region
$x \ge 3$ is a linear combination $f(x)=k_1 f_1(x)+k_3 f_3(x)$. The remaining
functions $u_1$ and $u_4$ are given by taking derivatives of $f(x)$.

In the region near the singular orbit the local deformation of
the ${\bf B}_{8}$
metric is characterized by two parameters, which we shall label as $m$ and
$q$~\cite{CGLP5}\cite{KY1}. The lower order terms near
the singular orbit are
\beqa
a(t) &=& m \left(1+\frac{3}{4}(t/m)^2 + \cdot \cdot \cdot \right)~, \CR
b_1(t) &=& b_2(t)=t \left(1-q(t/m)^2+ \cdot \cdot \cdot \right)~, \\
b_3(t) &=& t \left(1+(2q-1)(t/m)^2+ \cdot \cdot \cdot \right)~.
\nonumber
\eeqa
The ${\bf B}_{8}$ metric corresponds to $m=2\sqrt{2}\ell$
and $q=0$. As we have seen in Section 3, the $u$-system gives rise to
a diagonal deformation and the solutions $f(x)$ and (\ref{k0}) with $k_0=0$
induces the following transformation,
\beq
q=0 \rightarrow \tilde{q}=-5(k_1-28k_3)/32~, \quad
m=2\sqrt{2}\ell \rightarrow \tilde{m}
=2\sqrt{2}\ell \left(1+3(k_1-35k_3)/16 \right)~.
\eeq
The four-form $G^{SD}$ is not $L^2$-normalizable, unless we
do not impose the condition $k_3=0$ in the solution $f(x)$.
When $k_3=0$ the solution explicitly becomes
\beqa
u_{1} &=&-\frac{16k_1(5x^3-9x^2+15x-3)}{(x+1)^4(x-1)^3}~, \quad
u_{2}=u_{3}=\frac{32k_1 (x-3)}{(x+1)^4(x-1)^2}~, \CR
u_{4} &=& \frac{8k_1(x-3)(5x^2-2x+1)}{(x+1)^4(x-1)^3}~,
\eeqa
which gives an $L^2$-normalizable $G^{SD}$.
The $L^2$-normalizable
solution leads to a correlation in transformations of
the two parameters $m$ and $q$. However, we note that the
transformation of $m$ can be absorbed in an overall scaling of the metric
(a trivial deformation). Hence we conclude that the deformation of
${\bf B}_8$ that changes the asymptotic radius of the $S^1$ fiber
is essentially controlled by the $L^2$-normalizable $G^{SD}$.
We should mention that this $L^2$-normalizable solution was
obtained in \cite{CGLP3}, where it has been
noticed the ALC metric ${\bf B}_{8}$ admits a unique $L^2$-normalizable
harmonic four form of each duality. In \cite{CGLP3} the solution
with the opposite duality was used  to
construct supersymmetric brane solutions in $M$-theory.

The $\tilde{v}$-system (\ref{vsys}) can be solved easily and
we have regular solutions but growing exponentially at infinity.
Thus there are no $L^2$-normalizable solutions to the $\tilde{v}$-system
that might give deformations to off-diagonal metrics.
Though the lack of concrete examples of such metrics is a stumbling
block in taking further analyses, the fact that the diagonal deformation
arises from the $L^2$-normalizable solution suggests the corresponding
global deformation does not exist.

We now turn to the second example, which is given
by (\ref{SU3metric}) with
\beqa
a^2 &=& (r-m)(r+m/2)~, \quad b^2=(r+m)(r-m/2)~, \quad c_1^2 = c_2^2=r^2~, \CR
f^2 &=& \frac{9m^2(r+m)(r-m)}{16(r+m/2)(r-m/2)}~, \label{regular}
\eeqa
and $dt=\sqrt{r^2-m^2/4}/\sqrt{r^2-m^2}~dr$. The radial coordinate
$r$ runs from the singular orbit ${\bf CP}(2)$ at $r=m$ to infinity.
The ALC metric (\ref{regular}) is obtained with the subgroup $U(1)_{1,-1}$
and therefore within the generic case $(k \neq \ell)$ of $SU(3)/U(1)_{k,\ell}$
models. Remarkably the model with this choice of $U(1)$
subgroup has recently been employed to explore conifold like
geometric transitions in $Spin(7)$ geometry \cite{GST}.
The deformation admits only the $u$-system (\ref{usys}) with $k=-\ell=1$.
Using a variable $f=b^2 c^2 u_3-a^2 c^2 u_2$ and rescaling the radial
coordinate $x=r/m$, we obtain the following
differential equation of Fuchsian type~;
\beq
\frac{d^4f}{dx^4}+p_{1}(x)\frac{d^3f}{dx^3}+p_{2}(x)\frac{d^2f}{dx^2}
+p_{3}(x)\frac{df}{dx}+p_{4}(x)f=0~,
\eeq
where
\beqa
p_{1} &=&
\frac{256x^8-120x^6-118x^4+53x^2+7}{x(x^2-1)(x^2-1/4)(32x^4-12x^2-7)}~, \CR
p_{2} &=&
\frac{64x^8+296x^6-246x^4-61x^2-14}{x^2(x^2-1)(x^2-1/4)(32x^4-12x^2-7)}~, \CR
p_{3} &=&
-\frac{2(320x^8+64x^6+222x^4-8x^2-7)}{x^3(x^2-1)(x^2-1/4)(x^4-12x^2-7)}~, \CR
p_{4} &=&
\frac{156(8x^2+1)}{(x^2-1)(x^2-1/4)(32x^4-12x^2-7)}~.
\eeqa
The general solution is given by
\beqa
f_1 &=& \frac{x^2+x+1}{(x+1)(x^2-1/4)}~, \quad f_2 =
\frac{x^2(10x-7)}{10(x+1)(x^2-1/4)}~, \CR
f_3 &=& \frac{(x-1)^2(4x^4+12x^3+31x^2-9x-3)}{4(x+1)(x^2-1/4)}~, \quad
f_4 = \frac{1}{(x^2-1)(x^2-1/4)}~,
\eeqa
and so the regular solution in the region $x \ge 1$ is a linear combination
$f(x)=k_1 f_{1}(x)+k_{2}f_{2}(x)+k_{3}f_{3}(x)$. The $L^2$-normalizability
of $G^{SD}$ requires the condition for the coefficients, namely $k_2=10k_1$
and $k_3=0$. Finally we find
\beqa
u_1 &=& -\frac{k_1(x-1)(16x^2+9x-1)}{3x(x+1)^2(x^2+1/2)^2(x-1/2)^2}~, \quad
u_2 = -\frac{k_1(x-1)(14x^2+6x+1)}{6x^3(x+1)(x+1/2)^2(x-1/2)}~, \CR
u_3 &=& \frac{k_1(46x^4-14x^3+3x^2+2x-1)}{6x^3(x+1)^2(x+1/2)(x-1/2)^2}~,
\quad u_4 = -\frac{k_1(x-1)(3x+1)}{6x^3(x+1)^2(x+1/2)(x-1/2)}~. \label{solu}
\eeqa

The behavior of the metric on the $Spin^c$ bundle near the singular orbit
${\bf CP}(2)$ is given by \cite{CGLP5}\cite{KY1}
\beqa
a(t) &=& t \left(1-\frac{1}{2}(q+1)(t/m)^2+ \cdot \cdot \cdot \right)~, \CR
b(t) &=& m \left(1+\frac{1}{6}(4-n)(t/m)^2+ \cdot \cdot \cdot \right)~, \\
c(t) &=& m \left(1+\frac{1}{6}(5+n)(t/m)^2+ \cdot \cdot \cdot \right)~, \CR
f(t) &=& t \left(1+q(t/m)^2+ \cdot \cdot \cdot \right)~, \nonumber
\eeqa
where $q,~m$ are free parameters and the integer $n$ represents
an odd class in $H^2 ({\bf CP}(2), {\bf Z})$ that defines a $Spin^c$
structure of ${\bf CP}(2)$. The metric
(\ref{regular}) corresponds to $q=-26/27$ with $n=-1$ and the solution
(\ref{solu}) induces the transformation
\beq
q=-26/27 \rightarrow \tilde{q}=(-26+40 k_1)/27~, \quad
m \rightarrow \tilde{m}=m(1-k_1)~.
\eeq
\begin{flushleft}
{\bf 2. AC metric}
\end{flushleft}

Here for completeness we present a summary of
the results for AC metrics with special
holonomy. The qualitative feature of the deformation is different, since
we cannot find $L^2$-normalizable solutions.
There are three explicitly known AC metrics based on the coset space
$SU(3)/U(1)_{1,1}$ with the metric ansatz (\ref{SU3metric})~;

(a)~$Sp(2)$ metric on $T^*{\bf CP}(2)$ \cite{DS}\cite{CGLP2}
\beqa
a^2 &=&
\frac{1}{2}(r^2-m^2)~, \quad b^2=\frac{1}{2}(r^2+m^2)~, \quad
c_{1}^2 = c_{2}^2=r^2~, \CR
f^2 &=& \frac{r^2}{4}\left(1-(m/r)^4 \right) \quad \mbox{with}
\quad t = \int_{m}^{r} \frac{dr}{\sqrt{1-(m/r)^4}}~.
\eeqa

(b)~$SU(4)$ metric on the line bundle over $SU(3)/T^2$
\cite{PP}\cite{CGLP2}\cite{KY2}
\beqa
a^2 &=& b^2=\frac{r^2}{2}~, \quad c_{1}^2=c_{2}^2=r^2~, \CR
f^2 &=& \frac{r^2}{4}\left(1-(m/r)^8 \right) \quad \mbox{with}
\quad t= \int_{m}^{r} \frac{dr}{\sqrt{1-(m/r)^8}}~.
\eeqa

(c)~$Spin(7)$ metric on $T^{*}{\bf CP}(2)/{\bf Z}_2$ \cite{CGLP2}
\beqa
a^2 &=& b^2=\frac{9r^2}{10}~,
\quad c_{1}^2=c_{2}^2=\frac{9r^2}{25}\left(1-(m/r)^{10/3} \right)~, \CR
f^2 &=& \frac{9r^2}{100}\left(1-(m/r)^{10/3} \right) \quad \mbox{with}
\quad t= \int_{m}^{r} \frac{dr}{\sqrt{1-(m/r)^{10/3}}}~.
\eeqa

As before one can explicitly calculate the closed self-dual four forms
$G^{SD}$ for
these metrics and show that just as for the ALC metric there exists
a regular $G^{SD}$ which describes deformations at least locally.
However, the $L^2$-normalizability is lost for all
examples, which suggests that the deformations are non-dynamical.
   From the power series expansion we see that the corresponding deformation
would change the asymptotic behavior of the metric from AC type to ALC type.
This analysis seems consistent with what we have found
for the $L^2$-normalizability. We should remark that it is possible to
have an $L^2$-normalizable solution in the anti-self-dual side.
In fact this is the case for the Calabi metric
on $T^*{\bf CP}(2)$ \cite{CGLP2}.

In the following
we list the regular solutions to the $u$ and $v$-systems~;
\begin{flushleft}
(a-1)~$u$-system of $Sp(2)$ metric
\end{flushleft}
\beqa
u_1 &=& \frac{k_1 (x-1)(3x^2+9x+8)}{(x+1)^2}~, \CR
u_2 &=& \frac{k_1 (x-1)(3x^3+15x^2+x+1)}{2x^2(x+1)}+\frac{5k_2}{2x^2}~, \CR
u_3 &=& \frac{k_1 (x-1)^3(3x^2+6x+1)}{2x^2(x+1)^2}
-\frac{5k_2 (x+1)}{2x^2}~, \CR
u_4 &=& u_5 = -u_1/2 \quad \mbox{with} \quad x=(r/m)^2~.
\eeqa
\begin{flushleft}
(a-2)~$v$-system of $Sp(2)$ metric
\end{flushleft}
\beq
v_1=-v_3=\frac{k_3}{x}~, \quad v_2=-v_4=\frac{k_4}{x}~, \quad v_5=0~.
\eeq
\begin{flushleft}
(b-1)~$u$-system of $SU(4)$ metric
\end{flushleft}
\beqa
u_1 &=& \frac{k_1}{x^4}+\frac{k_3 w(x)}{3}~, \CR
u_2 &=& \frac{k_1}{x^4}+\frac{k_2}{x}-\frac{k_3 w(x)}{6}~, \CR
u_3 &=& \frac{k_1}{x^4}-\frac{k_2}{x}-\frac{k_3 w(x)}{6}~, \CR
u_4 &=& -u_5=\frac{k_3 \sqrt{x^4-1}}{8}~F[1/4,3/2,2;1-x^4] \label{sol1}
\eeqa
and
\beq
w = -x^2F[1/4,3/2,2;1-x^4]+\frac{3x^2(x^4-1)}{16}~F[5/4,5/2,3;1-x^4]~.
\eeq
Here $x=(r/m)^2$ and $F[\alpha,\beta,\gamma;x]$ denotes the hypergeometric
function.
\begin{flushleft}
(b-2)~$v$-system of $SU(4)$ metric
\end{flushleft}
There are no regular solutions.
\begin{flushleft}
(c-1)~$u$-system of $Spin(7)$ metric \cite{CLP}
\end{flushleft}
\beqa
u_1 &=& \frac{6k_1 z(z^{7/5}-1)}{z-1}~, \CR
u_2 &=& u_3 = k_1 w(z)+k_2z^{-4/5}(z-1)~, \CR
u_4 &=& k_1 w(z)-\frac{(k_2 + k_3)}{2}z^{-4/5}(z-1)~, \CR
u_5 &=& k_1 w(z)-\frac{(k_2 - k_3)}{2}z^{-4/5}(z-1) \label{sol2}
\eeqa
and
\beq
w=6+z^{7/5}-\frac{4z^{7/5}-11}{z-1}-\frac{5(z^{7/5}-1)}{(z-1)^2}~ \quad
\mbox{with} \quad z=(m/r)^{10/3}~.
\eeq
\begin{flushleft}
(c-2)~$v$-system of $Spin(7)$ metric
\end{flushleft}
\beq
v_{i}=K_{i}z^{-4/5}(z-1)~, \quad (i=1 \sim 5)
\eeq
where $K_i$ are constants satisfying $K_1=K_3$ and $K_2=K_4$.

Finally let us comment on the AC special holonomy metrics based on the coset
space $Sp(2)/Sp(1)$. Using the correspondence
between $Sp(2)/Sp(1)$ and $SU(3)/U(1)_{1,1}$
models, we can see the deformations of the AC metrics.
In fact the solution (\ref{sol1}) gives the deformation of
$SU(4)$ metric
on the line bundle over ${\bf CP}(3)$ \cite{PP}\cite{CGLP2},
and the solution (\ref{sol2})
the deformation of $Spin(7)$ metric on the bundle of chiral spinors
over $S^4$ \cite{BS}\cite{GPP}.



\vskip10mm

\begin{center}
{\bf Acknowledgements}
\end{center}

We would like to thank T. Eguchi, R. Goto and M. Naka for helpful discussions.
This work is supported in part by the Grant-in-Aid
for Scientific Research No. 14570073.


\section*{Appendix A}
\renewcommand{\theequation}{A.\arabic{equation}}\setcounter{equation}{0}
\begin{flushleft}
{\bf Convention of $SU(3)$ Maurer-Cartan forms}
\end{flushleft}

We use the following
$SU(3)$ Maurer-Cartan equation that is $\Sigma_3$ symmetric;
\beqa
d\sigma_1 &=& \Sigma_1 \wedge \tau_1 - \Sigma_2 \wedge \tau_2 + \kappa_A
T_A \wedge \sigma_2
+ \kappa_B T_B \wedge \sigma_2~, \CR
d\sigma_2 &=& - \Sigma_1 \wedge \tau_2 - \Sigma_2 \wedge \tau_1 - \kappa_A
T_A \wedge \sigma_1
- \kappa_B T_B \wedge \sigma_1~, \CR
d\Sigma_1 &=& \tau_1 \wedge \sigma_1 - \tau_2 \wedge \sigma_2 + \mu_A T_A
\wedge \Sigma_2
+ \mu_B T_B \wedge \Sigma_2~, \CR
d\Sigma_2 &=& - \tau_1 \wedge \sigma_2 - \tau_2 \wedge \sigma_1 - \mu_A T_A
\wedge \Sigma_1
- \mu_B T_B \wedge \Sigma_1~, \\
d\tau_1 &=& \sigma_1 \wedge \Sigma_1 - \sigma_2 \wedge \Sigma_2 + \nu_A T_A
\wedge \tau_2
+ \nu_B T_B \wedge \tau_2~, \CR
d\tau_2 &=& - \sigma_1 \wedge \Sigma_2 - \sigma_2 \wedge \Sigma_1 - \nu_A
T_A \wedge \tau_1
- \nu_B T_B \wedge \tau_1~, \CR
dT_A &=& 2\alpha_A \sigma_1 \wedge \sigma_2  + 2\beta_A \Sigma_1 \wedge
\Sigma_2
+ 2\gamma_A \tau_1 \wedge \tau_2~, \CR
dT_B &=& 2\alpha_B \sigma_1 \wedge \sigma_2  + 2\beta_B \Sigma_1 \wedge
\Sigma_2
+ 2\gamma_B \tau_1 \wedge \tau_2~. \nonumber
\eeqa
This form of the Maurer-Cartan equation is symmetric under the (cyclic)
permutation of $(\sigma_i, \Sigma_i,
\tau_i)$.
       From the Jacobi identity we see that
the parameters
$\alpha, \beta, \gamma, \kappa, \mu,\nu$, which describe the "coupling" of
the Cartan generators $\{ T_A, T_B\}$ satisfy
\beqa
& & \alpha_A + \beta_A + \gamma_A = 0~, \;
\alpha_B + \beta_B + \gamma_B = 0~,  \CR
\kappa_A &=& \frac{1}{\Delta}(\beta_B - \gamma_B), \;
\kappa_B = -\frac{1}{\Delta}(\beta_A - \gamma_A), \;
\mu_A = -\frac{1}{\Delta}(\alpha_B-\gamma_B), \\
\mu_B &=& \frac{1}{\Delta}(\alpha_A - \gamma_A), \;
\nu_A = \frac{1}{\Delta}(\alpha_B - \beta_B), \;
\nu_B = -\frac{1}{\Delta}(\alpha_A - \beta_A) \nonumber
\eeqa
with $\Delta=\beta_A \alpha_B -\alpha_A \beta_B$
leaving four free parameters $(\alpha_{A,B}, \beta_{A,B})$.
We may further put the "orthogonality" conditions;
\beqa
\alpha_A \alpha_B + \beta_A \beta_B + \gamma_A \gamma_B &=& 0~, \CR
\kappa_A \kappa_B + \mu_A \mu_B + \nu_A \nu_B &=& 0~,
\eeqa
which reduces one parameter.

\section*{Appendix B}
\renewcommand{\theequation}{B.\arabic{equation}}\setcounter{equation}{0}
\begin{flushleft}
{\bf Hitchin formulation of $Spin(7)$ manifolds}
\end{flushleft}

Recently Hitchin has shown that any hypersurface in a foliation of
eight-manifold with $Spin(7)$ holonomy carries a cosymplectic
$G_2$-structure \cite{Hitchin}. Here we describe the outline restricting to
$Spin(7)$ manifolds of cohomogeneity one.

Let $M=G/K$ be a seven dimensional homogeneous space with $G$-invariant
metric $\hat{g}$. Explicitly, using a basis of invariant 1-forms $E^i$,
we write
\beq
\hat{g}=g_{ij} E^i \otimes E^j~.
\eeq
A $G_2$-structure on $M$ is specified by fixing a three-form $\varphi$
which takes the form
\beq
\varphi=\frac{1}{3!}\varphi_{ijk} e^i \wedge e^j \wedge e^k. \label{g2}
\eeq
Here, $e^i \; (i=1 \sim 7)$ denote the vielbeins of $\hat{g}$, which
are identified with the generators of octonions obeying the relation
\beq
e^i e^j =\varphi_{ijk} e^k-\delta_{ij}.
\eeq
Furthermore (\ref{g2}) defines a cosymplectic $G_2$-structure on $M$ in case
$d*\varphi=0$.
Both weak $G_2$ and $G_2$ holonomy structures satisfy this condition and so
are special examples of cosymplectic $G_2$-structures. A classification of
compact homogeneous manifolds with weak $G_2$ holonomy is given
in \cite{FKMS} and the spaces $Sp(2)/Sp(1)$ and $SU(3)/U(1)_{k \ell}$
we discuss in this paper are in the members of this list.

Suppose $\rho_t=*\varphi(t)$ is a closed four-form in the space of
$G$-invariant
cosymplectic $G_2$-structures on $M$ for each $t \in I$(open interval).
If
\beq
\dot{\rho}_t+d*\rho_t=0~, \label{grad2}
\eeq
then the four-form
\beq
\Omega=dt \wedge *\rho_t - \rho_t
\eeq
gives a cohomogeneity-one metric with $Spin(7)$ holonomy on the
eight-manifols $N=I \times M$. In fact, we have
\beq
d\Omega=-dt \wedge d*\rho_t - dt \wedge \dot{\rho}_t=0~.
\eeq
The invariant metric $\hat{g}(t)$ on $M$ evolves via the equation (\ref
{grad2})
and induces the following $Spin(7)$ metric
\beqa
g &=& dt^2+\hat{g}(t) \CR
      &=& dt^2+g_{ij}(t) E^i \otimes E^j~.
\eeqa
Note that (\ref{grad2}) can be interpreted as the gradient flow equation
of the total volume
\beq
\sqrt{\mbox{det}g_{ij}(t)}\int E^1 \wedge E^2 \wedge \cdot \cdot \cdot
\wedge E^7~,
\eeq
when it is regarded as the functional of the closed form $\rho_t$~.

Conversely, if a $Spin(7)$ manifold is foliated by
homogeneous space $G/K$, we can write the $Spin(7)$ 4-form in the form
$dt \wedge *\rho_t - \rho_t$ with (\ref{grad2}) and $G/K$ carries an
invariant cosymplectic $G_2$-structure $\rho_t$ for each $t$.

\section*{Appendix C}
\renewcommand{\theequation}{C.\arabic{equation}}\setcounter{equation}{0}

\begin{flushleft}
{\bf Deformation of four dimensional hyperk\"ahler manifolds}
\end{flushleft}

A Riemannian manifold $(M,g)$ is called hyperk\"ahler when
it satisfies the following conditions~: (a) $M$ admits three
complex structures $J^a$ $(a=1,2,3)$ obeying the quaternionic
relations $J^a J^b =-\delta_{ab}-\epsilon_{abc} J^c$~; (b) the
metric $g$ is preserved by $J^a$~; (c) the two forms $\Omega^a$
defined by $\Omega^a(X,Y)=g(J^a X,Y)$ for all vector fields
$X,Y$ are three K\"ahler forms; $d\Omega^a =0$.
If the manifold is four dimensional,
the above condition is equivalent to
the (anti-)self-duality of the curvature two form.

In this appendix we discuss deformations of four dimensional
hyperk\"ahler manifold by a similar approach to $Spin(7)$ manifold
in this paper. Let us first recall a formulation
of the self-dual equations by Ashtekar.
In this approach hyperk\"ahler metrics are given by solutions
to the differential equations for volume-preserving vector fields.
The following proposition summarizes the result of \cite{AJS}\cite{MN}\cite{D}
relevant to our calculation.

\vspace{8mm}
{\bf Proposition} \quad Let $(M,\omega)$ be a four dimensional manifold
with a volume form $\omega$ and let $(V_0, V_1, V_2, V_3)$ be
volume-preserving vector fields forming an oriented basis for
$TM$ at each point.
Suppose in addition that the vector fields satisfy the equations
\beq
\bigl[V_0~,\; V_1\bigr]+\bigl[V_2~,\;V_3 \bigr]=0, \quad
\bigl[V_0~,\; V_2\bigr]+\bigl[V_3~,\;V_1 \bigr]=0, \quad
\bigl[V_0~,\; V_3\bigr]+\bigl[V_1~,\;V_2 \bigr]=0. \label{AS}
\eeq
Then the following metric $g$ is hyperk\"ahler
\beq
g=e^{\mu} \otimes e^{\mu}~, \quad e^{\mu}=\sqrt{\phi}\; V^{\mu}~,
\eeq
where $V^{\mu}$ is the dual one form of $V_{\mu}$ and
$\phi$ is a function defined by $\omega=\phi~ V^{0} \wedge
V^{1} \wedge V^2 \wedge V^3$, and the hyperk\"ahler
forms are given by the self-dual two forms~;
\beq
\Omega_{SD}^1=e^0 \wedge e^1~+e^2 \wedge e^3, \quad
\Omega_{SD}^2=e^0 \wedge e^2~+e^3 \wedge e^1, \quad
\Omega_{SD}^3=e^0 \wedge e^3~+e^1 \wedge e^2. \label{KA}
\eeq
We note that the duality of $\Omega_{SD}^i$ is fixed by
the orientation of the basis $V_\mu$ and the condition (\ref{AS}).

\vspace{8mm}

{\bf Remarks}\quad 1.  By using the orthonormal basis $e^{\mu}$
the spin connection $\omega_{\mu \nu}$ satisfies the equations
\beq
\omega_{01}+\omega_{23}=0~, \quad
\omega_{02}+\omega_{31}=0~, \quad
\omega_{03}+\omega_{12}=0~,
\eeq
and hence the corresponding curvature two form is automatically
anti-self-dual. The curvature two form has the reversed duality
to the hyperk\"ahler two forms.

2. The volume form of the metric is $\omega_g=\phi^2~ V^0 \wedge V^1
\wedge V^2 \wedge V^3$, which is different from the original
volume form $\omega$ one might expect.

\vspace{8mm}
Conversely, it is known that the volume-preserving vector fields
satisfying (\ref{AS}) can be locally constructed for any four
dimensional hyperk\"ahler manifold \cite{D}.

We now proceed to deformations of the hyperk\"ahler manifold $(M,g)$,
and use the orthonormal basis $e^{\mu}$ defined in the proposition.
Let $F_{ASD}$ be an anti-self-dual closed two form
(an ASD $U(1)$ instanton) on $M$~;
\beq
F_{ASD}=F_1(e^0 \wedge e^1-e^2 \wedge e^3)+F_2(e^0 \wedge e^2-e^3 \wedge e^1)
+F_3(e^0 \wedge e^3-e^1 \wedge e^2). \label{FASD}
\eeq
Let $\Omega_{SD}^a$ $(a=1,2,3)$ denote
the hyperk\"ahler forms given by (\ref{KA}).
We define deformed two forms by adding $F_{ASD}$ with
a small parameter $\epsilon$~;
\beq
\tilde{\Omega}^a=\Omega_{SD}^a+\epsilon~F_{ASD}~.
\eeq
Note that the shift is common to all the directions $(a=1,2,3)$.
When we introduce a new basis
\beq
\tilde{e}^{\mu}=e^{\mu}+ \frac{\epsilon}{2}~U^{\mu \nu}~e^{\nu}, \quad
(U^{\mu \nu}=U^{\nu \mu})
\label{UU}
\eeq
defined by
\beqa
U^{01} &=& F_2 -F_3, \quad U^{02} = -F_1+F_3, \quad
U^{03}=F_1 -F_2, \CR
U^{11} &=& F_1 -F_2 -F_3, \quad U^{22} = -F_1+F_2-F_3, \quad
U^{33}=-F_1 -F_2 +F_3, \CR
U^{12} &=& F_1 +F_2, \quad U^{13} = F_1+F_3, \quad
U^{23}=F_2 +F_3
\eeqa
with
\beq
U^{00}=-\sum_{i=1}^3~U^{ii} = F_1 + F_2 + F_3,
\eeq
it is easy to confirm that $\tilde{\Omega}^a$ takes the same
form as $\Omega_{SD}^a$ by means of the new basis $\tilde{e}^{\mu}$.
Thus the deformed metric $\tilde{g}=\tilde{e}^{\mu} \otimes
\tilde{e}^{\mu}$ is
hyperk\"ahler. Note that $F_{ASD}$ can be constructed from a
volume-preserving vector field $W=W_{\mu}V_{\mu}$ satisfying
the equation
\beq
\bigl[V_{\mu}~,\bigl[V_{\mu}~, W \bigr] \bigr]=0~.
\eeq
In fact the components of $F_{ASD}$ are given by \cite{OMYZ}
\beqa
F_1 = V_1(W_0)-V_0(W_1)+V_2(W_3)-V_3(W_2)~, \CR
F_2 = V_2(W_0)-V_0(W_2)+V_3(W_1)-V_1(W_3)~, \CR
F_3 = V_3(W_0)-V_0(W_3)+V_1(W_2)-V_2(W_1)~.
\eeqa

As an example let us consider the Atiyah-Hitchin metric \cite{AH}.
In \cite{Dan1}\cite{Dan2} it was shown that there exists a one-parameter
family of deformations of the metric by taking a hyperk\"ahler quotient of
a moduli space of $SU(3)$ monopoles.
By applying our method to this problem, the deformed
metric can be made more explicit, although the expression is
restricted to a small deformation from the Atiyah-Hitchin metric.

In the proposition, we take $M={\bf R} \times SO(3)$ and introduce
the left-invariant one forms $\sigma^i$ $(i=1,2,3)$ on $SO(3)$ given
by $A^{-1}dA=\sigma^i E_i$,~$A \in SO(3)$. Here, $E_i$ is the basis
of $so(3)$ algebra~,
\beq
E_1=\left(
\begin{array}{ccc}
0 & 0 & 0 \\
0 & 0 & 1 \\
0 & -1 & 0
\end{array}
\right),
E_2=\left(
\begin{array}{ccc}
0 & 0 & -1 \\
0 & 0 & 0 \\
1 & 0 & 0
\end{array}
\right),
E_3=\left(
\begin{array}{ccc}
0 & 1 & 0 \\
-1 & 0 & 0 \\
0 & 0 & 0
\end{array}
\right).
\eeq
Using Euler angles $\alpha, \beta$ and $\gamma$, we write
\beq
A=e^{\alpha E_1}e^{\beta E_2}e^{\gamma E_1} \label{SO}
\eeq
and then
\beqa
\sigma^1 &=& d\gamma+\cos\beta~d\alpha~, \CR
\sigma^2 &=& \sin\beta \cos\gamma~d\alpha-\sin\gamma~d\beta~, \CR
\sigma^3 &=& \sin\beta \sin\gamma~d\alpha+\cos\gamma~d\beta~.
\eeqa
Let us consider vector fields on $M$ \cite{Ab}
\beq
V_0 = \frac{\partial}{\partial t}~, \quad V_i =
A_{ij}\omega_{j}(t)\sigma_{j}~,
\eeq
where $A_{ij}$ represent the components of the matrix (\ref{SO}) and
$\sigma_j$ is the dual vector field of $\sigma^j$. The functions
$\omega_j(t)$ are
to be determined. The vector
fields $V_{\mu}$ preserve a volume form $\omega=dt \wedge \sigma^1 \wedge
\sigma^2 \wedge \sigma^3$. So applying the proposition to $V_{\mu}$, one can
obtain an SO(3)-invariant hyperk\"ahler metric of Bianchi IX type~;
\beq
g=\omega_{1}\omega_{2}\omega_{3}~dt^2+\frac{\omega_{2}\omega_{3}}{\omega_1}~
\sigma^1 \otimes \sigma^1+
\frac{\omega_{1}\omega_{3}}{\omega_2}~
\sigma^2 \otimes \sigma^2+
\frac{\omega_{1}\omega_{2}}{\omega_3}~
\sigma^3 \otimes \sigma^3~.
\eeq
and the condition (\ref{AS}) implies that $\omega_i$ must satisfy the equation
\beq
\dot{\omega_1}=\omega_2 \omega_3-\omega_1(\omega_2 + \omega_3) \label{elliptic}
\eeq
and its cyclic permutations. The flow equation (\ref{elliptic}) can be
solved in terms of elliptic
functions, which gives the Atiyah-Hitchin metric. An $SO(3)$-invariant
closed anti-self-dual two form $F_{ASD}$ on the Atiyah-Hitchin manifold
(sometimes called the Sen form)
was analysed by \cite{GR}\cite{Sen}, and it takes the form
\beq
F_{ASD}=f(t)(\omega_2 \omega_3~dt \wedge \sigma^1-\omega_1~\sigma^2
\wedge \sigma^3 ), \label{FA}
\eeq
where
\beq
f(t)=\exp\int_0^t(\omega_2+\omega_3-2 \omega_2 \omega_3/\omega_1)dt.
\eeq
The function $f$ has an exponential decline at $t \rightarrow \infty$
showing the $L^2$-normalizability of $F_{ASD}$.
If we rewrite (\ref{FA}) using the orthonormal basis
$e^{\mu}=\sqrt{\omega_1 \omega_2 \omega_3}~V^{\mu}$ defined by the
proposition, then the components $F_a$ $(a=1,2,3)$ of (\ref{FASD}) depend on
Euler angles through the functions $A_{ij}$. In fact
\beq
F_1=f(t)A_{11}~, \quad F_2=f(t)A_{21}~, \quad F_3=f(t)A_{31}~. \label{CO}
\eeq
This breaks the $SO(3)$ symmetry of the Atiyah-Hitchin metric
when the metric is deformed by $F_{ASD}$. Combining (\ref{CO})
with (\ref{UU}), we have an expression for the deformed metric
$\tilde{g}=g+\epsilon~h$~;
\beqa
h &=& (A_{11}+A_{21}+A_{31})f \left(\omega_1 \omega_2 \omega_3~dt^2
+\frac{\omega_2 \omega_3}{\omega_1}~\sigma^1 \otimes \sigma^1
-\frac{\omega_1 \omega_3}{\omega_2}~\sigma^2 \otimes \sigma^2
-\frac{\omega_1 \omega_2}{\omega_3}~\sigma^3 \otimes \sigma^3 \right) \CR
&+&(A_{12}+A_{22}+A_{32})f \left(\omega_1 \omega_2(dt \otimes \sigma^3+\sigma^3
\otimes dt) + \omega_3 (\sigma^1 \otimes \sigma^2
+\sigma^2 \otimes \sigma^1) \right) \CR
&+&(A_{13}+A_{23}+A_{33})f \left(-\omega_1 \omega_3(dt \otimes
\sigma^2+\sigma^2
\otimes dt)+\omega_2(\sigma^1 \otimes \sigma^3+\sigma^3 \otimes \sigma^1)
\right)~.
\eeqa
Let $K_i$~$(i=1,2,3)$ be vector fields on $SO(3)$ generated by the left
transformation $\exp(t E_i)$. Then the functions $A_{1j}+A_{2j}+A_{3j}$
vanish under the action $K=K_1+K_2+K_3$. Thus the metric $\tilde{g}$ has
an isometry $SO(2)$ generated by $K$ as has been claimed in \cite{Dan2}.




\end{document}